\documentclass[twocolumn,amsmath,amssymb,prd,preprintnumbers,superscriptaddress,showpacs]{revtex4}

\usepackage{graphicx}
\usepackage{latexsym}
\usepackage{bm}
\usepackage{amsmath}
\usepackage{amssymb}

\def\ktr{\tilde{\kappa}_{\rm tr}}
\def\kem{\tilde{\kappa}_{e-}}
\def\kep{\tilde{\kappa}_{e+}}
\def\kop{\tilde{\kappa}_{o+}}
\def\kom{\tilde{\kappa}_{o-}}

\def\fr#1#2{{{#1} \over {#2}}}
\def\half{{\textstyle{1\over 2}}}
\def\quar{{\textstyle{1\over 4}}}

\def\lsim{\mathrel{\rlap{\lower3pt\hbox{$\sim$}}
    \raise2pt\hbox{$<$}}}
\def\gsim{\mathrel{\rlap{\lower3pt\hbox{$\sim$}}
    \raise2pt\hbox{$>$}}}
\def\sqr#1#2{{\vcenter{\vbox{\hrule height.#2pt
         \hbox{\vrule width.#2pt height#1pt \kern#1pt
         \vrule width.#2pt}
         \hrule height.#2pt}}}}

\def\prt{\partial}
\def\lrpartial{\raise 1pt\hbox{$\stackrel\leftrightarrow\partial$}}

\def\lrDmu{{\hskip -3 pt}\stackrel{\hskip -1.5 pt\leftrightarrow}{D_\mu}{\hskip -2pt}}

\newcommand{\bit}{\begin{itemize}}
\newcommand{\eit}{\end{itemize}}

\newcommand{\beq}[1]{\begin{equation}\label{#1}}
\newcommand{\eeq}{\end{equation}}
\newcommand{\bea}[1]{\begin{eqnarray}\label{#1}}
\newcommand{\eea}{\end{eqnarray}}
\newcommand{\ba}{\begin{array}}
\newcommand{\ea}{\end{array}}

\newcommand{\rf}[1]{(\ref{#1})}

\def\fr#1#2{{{#1} \over {#2}}}
\def\etal{{\it et al.}}

\begin{document}

\title{Limits on isotropic Lorentz violation in QED from collider physics}

\author{Michael A.~Hohensee}
\affiliation{Department of Physics, Harvard University, Cambridge, Massachusetts, 02138, USA}
\author{Ralf Lehnert}
\affiliation{Max--Planck--Institut f\"ur Physik,
F\"ohringer Ring 6, 80805 M\"unchen, Germany}
\affiliation{Instituto de Ciencias Nucleares,
Universidad Nacional Aut\'onoma de M\'exico,
A.~Postal 70-543, 04510 M\'exico D.F., Mexico}
\author{David F.~Phillips}
\affiliation{Harvard--Smithsonian Center for Astrophysics, Cambridge, Massachusetts 02138, USA}
\author{Ronald L.~Walsworth}
\affiliation{Department of Physics, Harvard University, Cambridge, Massachusetts, 02138, USA}
\affiliation{Harvard--Smithsonian Center for Astrophysics, Cambridge, Massachusetts 02138, USA}

\date{\today}

\begin{abstract} 
  We consider the possibility that Lorentz violation can generate differences
  between the limiting velocities of light and charged matter.  Such effects would lead to efficient vacuum
  Cherenkov radiation or rapid photon decay.  The absence of such
  effects for $104.5\;$GeV electrons at the Large Electron Positron
  collider and for $300\;$GeV photons at the Tevatron therefore
  constrains this type of Lorentz breakdown.  Within the context of
  the standard-model extension, these ideas imply an
  experimental bound at the level of
  $-5.8\times10^{-12}\leq\ktr-(4/3)c_e^{00}\leq1.2\times 10^{-11}$ tightening
  existing laboratory measurements by 3--4 orders of magnitude.
  Prospects for further improvements with terrestrial and
  astrophysical methods are discussed.
\end{abstract}


\maketitle

\section{Introduction}
\label{introduction}

Established physics is successfully described by two distinct
theories: general relativity (GR) and the standard model (SM) of
particle physics.  These two theories are commonly believed to arise
as the low-energy limit of a more fundamental Planck-scale framework
that consistently merges gravity and quantum mechanics.  Since direct
measurements at this scale are presently impractical, experimental
research in this field relies largely on ultrahigh-precision searches
for Planck-suppressed effects at attainable energies.

One candidate effect within this context is a minute breakdown of
Lorentz invariance \cite{cpt07}.  Lorentz symmetry represents a cornerstone of both GR and the
SM, so that any observed deviation from this symmetry would imply new
physics.  A number of theoretical approaches to
underlying physics, such as strings \cite{ksp}, noncommutative field
theories \cite{ncqed}, cosmologically varying fields
\cite{spacetimevarying}, quantum gravity \cite{qg}, random-dynamics
models \cite{fn}, multiverses \cite{bj}, brane-world scenarios
\cite{brane}, and massive gravity \cite{modgrav}, are known to
accommodate small violations of Lorentz invariance at low energies.
Searches for such violations are also motivated by the seemingly fundamental character of Lorentz symmetry: it should be buttressed as
firmly as possible by experimental evidence.

At currently attainable energies, Lorentz-violating effects are
expected to be described by an effective field theory \cite{kp}.  The
standard-model extension (SME) provides the general framework in this
context \cite{sme,ssb}, containing both GR and the SM as limiting
cases.  The additional Lagrangian terms of the SME include all
operators for Lorentz violation that are scalars under coordinate
changes.  The SME has already provided the basis for the analysis of
numerous experimental searches for Lorentz breakdown \cite{kr},
including ones with photons \cite{km0102,photonexpt,future,kappatr},
electrons \cite{eexpt,astro,eexpt2and3}, protons and neutrons
\cite{ccexpt,clock,wolf06,spaceexpt,bnsyn}, mesons \cite{hadronexpt}, muons
\cite{muexpt}, neutrinos \cite{nuexpt}, the Higgs \cite{higgs}, and
gravity \cite{gravexpt,bak06}.

The speed of light in particular has played a key role in both the
conception of Lorentz symmetry and its early experimental tests.  The
continuing importance of electrodynamics to the subject is illustrated
by the high precision with which the SME's photon sector is bounded.
However, the most notable exception to these tight limits has been the
SME $\ktr$ coefficient, which parameterizes isotropic shifts in the speed of
light.  While the gap in precision to other laboratory constraints on
electrodynamics 
was, until recently, at least 5 orders of
magnitude, we recently published new results employing data from
CERN's Large Electron Positron (LEP) collider and Fermilab's
Tevatron, improving existing laboratory limits on $\ktr$ by 3--4
orders of magnitude~\cite{hohensee09}.

In this paper, 
we expand on our recent work 
to provide a detailed analysis of the sensitivity of
collider experiments to $\ktr$. 
The outline of this paper is as
follows.  In Sec.~\ref{review}, we review the basics of
Lorentz-violating electrodynamics coupled to charged matter and
present the basic idea behind the physics leading to our constraints
on $\ktr$.  Section~\ref{vcr} employs the absence of vacuum Cherenkov
radiation to limit positive values of $\ktr$.  Constraints on negative
values of $\ktr$, inferred from photon stability, are derived in
Sec.~\ref{decay}.  Section~\ref{sum} presents a brief summary and
mentions future possibilities for measuring $\ktr$, along with some of
the associated projected bounds.  Supplementary material is collected
in two appendices.  Unless noted otherwise, we work in natural units
$c=\hbar=1$, and our convention for the metric signature is
$(+,-,-,-)$.

\section{Basics}
\label{review}

The  photon and electron sectors 
 of the minimal SME \cite{sme} are described by the Lagrangian
\bea{lag1}
\mathcal{L} \!& = &\! {}-\quar F^{2}-\quar(k_{F})^{\kappa\lambda\mu\nu}F_{\kappa\lambda}F_{\mu\nu}+(k_{AF})^{\mu}A^{\nu}\tilde{F}_{\mu\nu}\nonumber\\
&&\! {}+\half{\it i}\,\overline{\psi}\,
{\Gamma}^{\nu}
\!\!\stackrel{\;\leftrightarrow}
{D}_{\nu}\! {\psi}
-\overline{\psi}M{\psi}\;,
\eea
where
\bea{defs}
{\Gamma}^{\nu} \! & \equiv & \! {\gamma}^{\nu}+c_e^{\mu \nu}
{\gamma}_{\mu}+d_e^{\mu \nu}{\gamma}_{5}
{\gamma}_{\mu}
\; ,\nonumber\\
M \! & \equiv &  \! m_e
+b_{e}^{\mu}{\gamma}_{5}
{\gamma}_{\mu}+\half H_e^{\mu \nu}
{\sigma}_{\mu \nu}
\; ,
\eea
$F^{\mu\nu}=\prt_\mu A_\nu-\prt_\nu A_\mu$ is the
electromagnetic field-strength tensor, and $\tilde{F}^{\mu\nu}=(1/2)
\epsilon^{\mu\nu\rho\sigma}F_{\rho\sigma}$ denotes its dual.  The
spinor $\psi$ describes electrons and positrons of mass $m_e$, and the usual
U(1)-covariant derivative is denoted by $D^\mu=\partial^\mu+i e
A^\mu$.  The SME coefficients
$(k_{F})^{\mu\nu\rho\lambda}$, $(k_{AF})^{\mu}$, $b_e^\mu$,
$c_e^{\mu\nu}$, $d_e^{\mu\nu}$, and $H_e^{\mu\nu}$ control the extent of
Lorentz and CPT violation.

In this work, we are primarily interested in the $\ktr$ component
of $(k_{F})^{\kappa\lambda\mu\nu}$, and will set
$(k_{AF})^{\mu}$ to zero.
The $(k_{F})^{\kappa\lambda\mu\nu}$ coefficient possesses the
symmetries of the Riemann curvature tensor, and its double trace
vanishes $(k_{F})^{\kappa\lambda}{}_{\kappa\lambda}=0$, leaving 19
independent components.
To exhibit $\ktr$, 
we decompose $(k_{F})^{\kappa\lambda\mu\nu}$ 
such that the electromagnetic component of the Lagrangian \rf{lag1} becomes:
\begin{multline}
\label{lagr2}
\!\!\!\!\!
{\cal L}=\frac{1}{2}\left[(1+\ktr)\mathbf{E}^{2}-(1-\ktr)\mathbf{B}^{2}\right]+\frac{1}{2}\mathbf{E}\cdot(\kep+\kem)\cdot\mathbf{E}\\
-\frac{1}{2}\mathbf{B}\cdot(\kep-\kem)\cdot\mathbf{B}+\mathbf{E}\cdot(\kop+\kom)\cdot\mathbf{B}\;.
\end{multline}
Here, the dimensionless parameter $\ktr$ and the dimensionless and
traceless $3\times 3$ matrices $\kem$, $\kep$, $\kop$, and $\kom$ are
defined in terms of the $(k_{F})^{\kappa\lambda\mu\nu}$ coefficients
\cite{km0102} with
$\ktr\equiv (2/3)(k_{F})_\mu{}^{0\mu0}$.  Note that the above
decomposition of $(k_{F})^{\kappa\lambda\mu\nu}$ into $\tilde{\kappa}$
coefficients is not manifestly coordinate independent: under changes
of the observer inertial frame, the various $\tilde{\kappa}$
parameters mix.  To facilitate comparisons between different
experimental tests, a reference coordinate system must therefore be
selected.  A conventional choice is the Sun-centered celestial
equatorial frame \cite{km0102}.

The parameterization determined by Eq.~\rf{lagr2}
is particularly intuitive 
because of its analogy to conventional electrodynamics 
in macroscopic media \cite{sme,km0102}. 
For example, 
all $\tilde{\kappa}$ coefficients modify the photon dispersion relation, 
and thus the phase speed of light $c_{\rm ph}$. 
A subset of them, 
namely $\kom$ and $\kep$, 
affect each of the two electromagnetic-wave polarizations differently 
leading to birefringence. 
The absence of this type of birefringence 
in spectropolarimetric studies of cosmological sources
constrains $\kom$ and $\kep$ at the level of $10^{-37}$~\cite{km06}. 
The remaining coefficients $\kem$, $\kop$, and $\ktr$ 
arise from the 
\beq{ktildedef} 
\tilde{k}^{\mu\nu}\equiv (k_F)_\alpha{}^{\mu\alpha\nu}
\eeq 
component.  They lead to polarization-independent shifts in $c_{\rm ph}$,
so that other types of measurements are necessary.  A particularly
sensitive measurement involves optical- or microwave-cavity
experiments that search for parity-even anisotropies in $c_{\rm ph}$.
Assuming negligible Lorentz violation effects on the cavity itself,
these tests set limits on the elements of $\kem$ at the level of
$10^{-17}$ \cite{her07}.  The parity-odd matrix $\kop$ and the
isotropic $\ktr$ can be constrained indirectly with such measurements
via higher-order effects: the laboratory is boosted with respect to
the Sun-centered celestial equatorial frame, so that $\kem$, $\kop$,
and $\ktr$ mix.  The absence of any observed resultant effect upon the cavity resonances yields $\kop\lsim10^{-13}$~\cite{her07}, and
$|\ktr|<1.8\times 10^{-8}$~\cite{secondorderres}.  
The latter had represented 
the best laboratory bound on~$\ktr$
prior to the recent paper~\cite{hohensee09}
expanded here.

\begin{figure}
\vskip10pt
\begin{center}
\includegraphics[width=0.80\hsize]{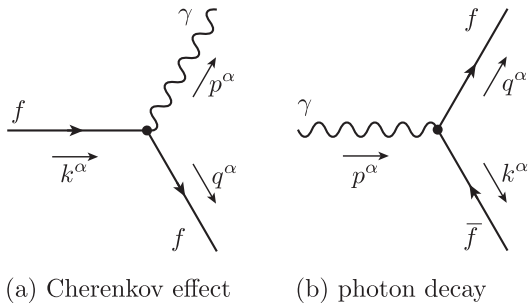}
\end{center}
\vskip-10pt
\caption{Two possible electromagnetic-vertex configurations. 
Charged fermions and antifermions are denoted by 
$f$ and $\overline{\hspace{-.5mm}f\hspace{.5mm}}$, 
respectively.
The photon is labeled by $\gamma$.
In the conventional case, 
energy--momentum conservation 
together with ordinary dispersion relations
prohibits the three external legs 
from going simultaneously on-shell. 
In the presence of 
the Lorentz-violating $\ktr$ modification 
of the photon dispersion relation
kinematically allows 
reactions (a) and (b) 
for $\ktr>0$ and $\ktr<0$, 
respectively.
}
\label{fig0} 
\end{figure}

The present study exploits direct physical effects of $\ktr$ 
to obtain improved constraints on this coefficient. 
The basic idea is 
that dispersion-relation modifications due to $\ktr$ 
would not only change $c_{\rm ph}$, 
but also affect the kinematics of the electromagnetic vertex. 
In contrast to the conventional case, 
the three external legs of the vertex 
can go simultaneously on shell 
allowing various particle reactions to proceed
that are normally forbidden by Lorentz symmetry. 
We focus on two such reactions, 
each occurring only for a specific sign of  $\ktr$. 
The first is vacuum Cherenkov radiation
\beq{vcrrec}
f \to f +\gamma\quad\textrm{for}\quad\ktr>0\;,
\eeq
and the second is photon decay 
\beq{phdec}
\gamma \to f +\overline{\hspace{-.5mm}f\hspace{.5mm}}\quad\textrm{for}\quad\ktr<0\;.
\eeq
Here, $\gamma$ denotes a photon, 
$f$ a charged fermion, 
and $\overline{\hspace{-.5mm}f\hspace{.5mm}}$ 
the corresponding antifermion. 
These processes are depicted in Fig.~\ref{fig0}.

In what follows, we will explore how the observed absence of the
reactions~\rf{vcrrec} and~\rf{phdec} provides bounds on
$\ktr$, expanding on our recently published paper on this subject~\cite{hohensee09}.  Similar ideas have been exploited
previously~\cite{cg99,kappatr}, primarily in the context of purely
kinematical dispersion-relation tests~\cite{previous}.  In the present
case, the underlying SME Lagrangian permits the inclusion of dynamical
features, such as the rate at which the reactions~\rf{vcrrec}
and~\rf{phdec} proceed.  Dynamical considerations are often necessary
to obtain convincing and conservative results~\cite{thres,vcr}.

In addition to $\ktr$, 
other Lorentz-violating SME coefficients 
can lead to dispersion-relation modifications, 
and thus 
to the reactions~\rf{vcrrec} or~\rf{phdec}.
However, 
the effects of such additional coefficients 
can be safely neglected 
as long as their scale ${\cal S}$ 
is small compared to 
the $\ktr$ constraint to be determined. 
The other relevant coefficients 
are those of the fermion $f$, 
the remaining $k_F$ components 
(i.e., the $\tilde{\kappa}$ matrices), 
and the $k_{AF}$ coefficient in Lagrangian~\rf{lag1}.
Since we will primarily consider $f$ to be an electron, 
the relevant matter-sector coefficients 
in the minimal SME
are $b_e^{\mu}$,  
$c_e^{\mu\nu}$, 
$d_e^{\mu\nu}$, 
and 
$H_e^{\mu\nu}$ 
of the electron. 
With these considerations, 
the scale ${\cal S}$ is given by
\beq{scale}
{\cal S}\equiv\textrm{max}\left(\tilde{\kappa}_{e\pm}, \tilde{\kappa}_{o\pm}, \frac{k_{AF}}{m_e},
\frac{b_e}{m_e},c_e,d_e,\frac{H_e}{m_e}\right)\,.
\eeq
Here, 
the absolute values of the individual components 
of the SME coefficients are implied, 
and $m_e$ the electron mass, 
as before. 

To interpret and determine the scale ${\cal S}$ correctly, 
certain subtleties need to be taken into account.  
One such is 
that elements of the electron's $c_e^{\mu\nu}$ coefficient 
are physically equivalent to
the nonbirefringent $\tilde{\kappa}$ parameters, 
as can be established 
by a coordinate redefinition.  
This means 
that only the anisotropic components of $c_e^{\mu\nu}$ 
should enter the determination of ${\cal S}$
because only anisotropic $\tilde{\kappa}$'s 
occur in ${\cal S}$.
More importantly, 
the equivalence of $c_e^{\mu\nu}$ 
and the nonbirefringent $\tilde{\kappa}$'s implies
that our ultimate constraint upon $\ktr$ is, 
strictly speaking, 
a constraint upon the linear combination $\ktr-(4/3)c_e^{00}$.
In what follows,
we will often scale the coordinates 
such that $c_e^{00}=0$, 
but undo this special choice of scaling 
and reinstate $c_e^{00}$ 
when stating results.
A more complete discussion of the above issues 
is contained in Appendix~\ref{rescaling}.  
The result of interest in the present context is 
that currently ${\cal S}\sim10^{-13}$ 
dominated by the $\kop$ matrix coefficient~\cite{kr}.  
So we may indeed focus on $\ktr$ 
and ignore other types of Lorentz violation 
for photon--electron interactions.

We note in passing that 
vacuum Cherenkov radiation can also occur for antifermions 
and that further unconventional processes,
such as fermion--antifermion annihilation into a single photon, 
are possible.
Moreover,
two-photon emission and absorption processes, 
synchrotron radiation, and inverse Compton processes 
can also be modified. 
Some of these effects have been employed in astrophysical contexts 
to extract general bounds on $\Delta c_{\rm ph}/c_{\rm ph}$ down to the $10^{-16}$ level~\cite{astro}.

\section{Vacuum Cherenkov radiation}
\label{vcr}

In the present context, 
the vacuum Cherenkov effect~\rf{vcrrec}
can only occur 
for positive $\ktr$. 
To leading order, the modified dispersion relation for a photon with wave vector $p^{\mu}\equiv(E_{\gamma},\vec{p})$ is~\cite{km0102}
\beq{disprel}
E_{\gamma}^2-(1-\ktr)\vec{p}^{\,2}=0\;.
\eeq
We scale the coordinates such that the fermion dispersion relation
remains conventional.  Energy-momentum conservation for the
process~\rf{vcrrec} then yields a threshold energy $E_{\rm VCR}$,
\beq{vcrenergy}
E_{\rm VCR}=\frac{1-\ktr}{\sqrt{(2-\ktr)\ktr}}m
=\frac{1}{\sqrt{2\ktr}}m+{\cal O}\left(\sqrt{\ktr}\right)\;,
\eeq
which corresponds to the kinetic energy of a fermion with mass $m$ 
moving as fast as photons obeying~\rf{disprel} 
in the vacuum \cite{ba08}.
For charges with energies above $E_{\rm VCR}$, 
vacuum Cherenkov radiation 
is kinematically allowed.

Here we constrain positive values of $\ktr$ from the observed absence of vacuum
Cherenkov radiation in nature.  To set such constraints, vacuum
Cherenkov radiation would have to be efficient enough to be
observable.  Close to the threshold energy $E_{\rm VCR}$, the dominant
process is single-photon emission, such that the charge falls below
threshold; an estimate for the corresponding rate is~\cite{ba08}:
\beq{vcrate}
\Gamma_{\rm VCR}=\alpha\, Z ^{2} m^{2}\,\frac{(E_f-E_{\rm VCR})^{2}}{2E_f^{3}}\;,
\eeq
where $\alpha$ is the fine-structure constant, 
$Z$ the charge 
measured in multiples of the elementary charge, 
and $E_f$ the fermion energy.
This shows 
the effect is undoubtedly efficient: 
for example, 
a $104.5$ GeV electron 
with an energy of 1\% above the threshold~\rf{vcrenergy}
would reach subluminal speeds 
after traveling an average distance of $23\,$cm. 
We therefore conclude 
that limits on $\ktr$ 
can indeed be established 
from the observed absence at particle colliders of the vacuum Cherenkov effect
for low-mass charges 
at the highest possible energies.  

\subsection{Bounds from collider experiments}
\label{collider}
Common to all analyses of collider experiments is a precise knowledge of the species and energy of the potential vacuum Cherenkov emitter.  At present, the LEP experiment provides the best compromise between a charge's mass vs.\ its energy for terrestrial Cherenkov constraints.  As a result, we can immediately determine that measurements of $E_{\rm VCR}$ derived from the LEP $e^{+}e^{-}$ beams will constrain $\tilde{k}^{\mu\nu}-2c_{e}^{\mu\nu}$, independent of Lorentz-violating effects for other particles.  As shown in Appendix~\ref{rescaling}, for the energies attained at LEP ($\sim 100\:$~GeV), $\ktr-(4/3)c_{e}^{00}$ is the only SME coefficient combination that can contribute to vacuum Cherenkov radiation, permitting a significantly simplified study.  An analysis of LEP data has the potential to yield rigorous one-sided improvements upon previous laboratory constraints on $\ktr$.

The LEP collider
was a circular particle accelerator 
approximately $27\,$km in circumference.  
This accelerator was an exquisitely precise and carefully controlled device
with a relative uncertainty in the center-of-mass energy 
$\Delta E_{\rm CM}/E_{\rm CM}$ 
less than $2.0\times 10^{-4}$ \cite{LEP05}.  
To keep the uncertainty at this level, 
minute effects
such as Earth tides, 
variations in the pressure of the local water table,  
and even seasonal variations in the volume of the nearby lake 
needed to be taken into account \cite{LEP05}.
The highest laboratory-frame energy attained at LEP 
was $E_{\rm LEP}=104.5\,$GeV.  
We can obtain a first estimate for a limit on $\ktr$ 
by arbitrarily setting $E_{\rm VCR}=100\,$GeV.
Then,  
$104.5\,$GeV electrons or positrons would fall below threshold 
after traveling approximately $1.2\,$cm. 
This length is far shorter
than the distance between superconducting radio-frequency cavities at LEP
or even the dimensions ($5.8\,$m) of each of the dipole bending magnets \cite{LEP05}, 
so that such an effect would have been readily apparent.  
The observed absence of such effects at LEP implies $E_{\rm VCR}>100\,$GeV;
together with Eq.~\rf{vcrenergy}, we then obtain $0\leq\ktr \leq 1.3\times 10^{-11}$.
A more refined line of reasoning is presented 
in the next paragraphs.

At $E_{\rm LEP}=104.5\,$GeV, 
the energy loss due to conventional synchrotron radiation 
was $U_0=3.486\,$GeV per electron or positron per turn~\cite{LEP05}. 
The LEP circumference of $26\,659\,$m 
then predicts an average energy loss per distance travelled of 
\beq{synchloss}
\frac{dE_{\rm syn}}{dL}=2.580\times10^{-20}\,{\rm GeV}^2\,.
\eeq
One of the three energy-calibration methods at LEP 
relied upon the dependence of the synchrotron tune 
on the energy loss. 
For this reason, 
a precise determination of the energy loss was paramount. 
Deviations from the value~\rf{synchloss} 
arise through parasitic-mode losses, 
finite beam size and other quadrupole effects, 
and losses in the corrector dipoles. 
The sum of these contributions
is conservatively estimated
to be $0.5\,$MeV per turn per particle 
with at most a 20\% uncertainty~\cite{LEP05}.
This implies that
\beq{cherloss}
\frac{dE_{\rm Cher}}{dL} \leq 10^{-4}\,\frac{dE_{\rm syn}}{dL}\;,
\eeq
where $dE_{\rm Cher}/dL$ 
denotes the energy loss per distance 
due to vacuum Cherenkov radiation.

The final step is 
to determine a lower bound for $E_{\rm VCR}$ 
such that the inequality~\rf{cherloss} 
together with the value~\rf{synchloss} 
is satisfied. 
To this end, 
recall that for charges near $E_{\rm VCR}$ 
the dominant Cherenkov process 
for reaching subthreshold energies 
proceeds via single-photon emission~\rf{vcrate}. 
The energy loss per Cherenkov event 
must therefore be greater than $E-E_{\rm VCR}$. 
The average distance $L$ traversed 
by an electron before Cherenkov emission occurs 
is $1/\Gamma_{\rm VCR}$. 
With Eq.~\rf{vcrate}, 
this yields 
\beq{cherlosstheo}
\frac{dE_{\rm Cher}}{dL} \geq \alpha\, m_e^2\,
\frac{(E_{\rm LEP}-E_{\rm VCR})^3}{2E_{\rm LEP}^3}\;,
\eeq
where $m_e=5.11\times10^{-4}\,$GeV 
denotes the electron mass, 
as before. 
It follows 
that $E_{\rm VCR}$ can at most be $1.5\,$MeV 
below $E_{\rm LEP}=104.5\,$GeV.  
With Eq.~\rf{vcrenergy}, we then obtain
\beq{VCRbound}
0\leq\ktr-\tfrac{4}{3}c_{e}^{00}\leq1.2\times10^{-11}\;,
\eeq
explicitly including the contribution of $c_{e}^{00}$.
The above reasoning also shows 
that the uncertainty in the bound~\rf{VCRbound} 
is primarily determined by 
the accuracy of the electron-energy measurement.
As this limit is still much larger than the scale ${\cal S}$ 
defined in Eq.~\rf{scale}, 
other photon- or electron-sector coefficients
are not further constrained by this reasoning. 
At the same time, 
this provides the justification 
for dropping these additional coefficients 
from our analysis.

\subsection{Cosmic-ray analyses}

Vacuum Cherenkov tests
compare the respective group velocities $\vec{v}_X$ and $\vec{v}_\gamma$ 
of the charge $X$ and the photon $\gamma$.
Since both particles 
may exhibit independent Lorentz-violating effects,
vacuum Cherenkov radiation will typically depend on
parameters originating from {\em both} the photon 
{\em and} the charge sectors of the SME.  This means that any analysis of Lorentz-symmetry violation based on vacuum Cherenkov physics must address the following points:
First, 
in the absence of independent constraints from other experiments, 
a Cherenkov analysis 
must incorporate all relevant Lorentz-symmetry violating coefficients 
from the photon as well as the charge.
Second, 
the nature of the charge must be known, since it would otherwise be unclear as to which SME coefficients are actually constrained.
The analysis of the previous subsection easily addresses the second point, as it involves the electrons and positrons at LEP.  The first point is also addressed in detail in Appendix~\ref{rescaling}: 
the availability of complementary experimental results leads to an estimation of the scale ${\cal S}$ in Eq.~\ref{scale} which 
justifies dropping all but the single combination $\ktr-(4/3)c_e^{00}$
from consideration. 

Charged ultrahigh-energy cosmic rays (UHECRs) 
offer the potential 
to yield the tightest limits on positive values of $\ktr$, 
as they possess energies 
orders of magnitude above those available in any laboratory~\cite{kappatr}.  
Unfortunately, 
efforts to use observations of UHECRs to constrain Lorentz violation in the photon sector are currently beleaguered by a number of interpretational difficulties.
Chief among them is the lack of certainty 
as to the composition of UHECR primaries, 
leading to an associated uncertainty 
as to which SME coefficients are constrained. 
Although the observed UHECR primaries are believed to be single protons, 
the possibility 
that the observations could be due to the scattering of more massive nuclei, 
high-energy photons, 
or Lorentz-violating particles exhibiting no or a qualitatively different Cherenkov effect, 
such as stable neutral pions or neutrons~\cite{cg99}, 
cannot yet be excluded.  
This uncertainty will likely be ameliorated 
in coming years with continued observations.

If the UHECR primaries are found to be single protons,
then in principle the analysis in Ref.~\cite{kappatr} establishes 
that constraints on a combination of photon and proton 
SME coefficients at the $10^{-21}\ldots 10^{-22}$ level
can be obtained.
In contrast to our LEP-electron study, however, 
there is an insufficient number 
of complementary experimental bounds 
to solely focus on the nine parameters 
$2c_{p}^{\mu\nu}-\tilde{k}^{\mu\nu}$~\cite{fn0,fn0.5}.  
For example, 
the $d^{\mu\nu}_p$ SME coefficient 
also affects the maximal attainable velocity of the proton 
and can therefore lead to vacuum Cherenkov radiation.
But thus far, 
only two of its nine components have been bounded~\cite{clock}, 
so in general this coefficient needs to be taken into consideration
when performing vacuum Cherenkov tests with protons.

If, instead, 
the UHECR primaries are identified as atomic nuclei 
such as He, $^{12}$C, or even $^{56}$Fe nuclei, 
as is assumed in Ref.~\cite{kappatr}, 
various additional considerations are necessary.
For example,
the physical system under consideration now 
also contains neutrons 
in addition to photons and protons.
This means 
that in principle 
neutron SME coefficients 
need to be considered as well. 
Moreover, 
the potential Cherenkov emitter
is now a bound state without its own SME coefficients.
One could introduce effective Lorentz-violating parameters
for the nucleus, 
but this would hamper comparison 
with other limits on Lorentz violation 
for the photon, proton, or neutron. 
It is therefore preferable 
to determine the group velocity of the nucleus
in terms of the proton's and neutron's 
SME coefficients 
via simplified nuclear modeling. 
To this end, 
it may be possible to employ 
the nuclear Schmidt model 
along the lines of the analysis in Ref.~\cite{clock}.

\section{Photon decay}
\label{decay}

For negative $\ktr$, 
the phase speed of light is greater than unity~\cite{fn1}.  
Vacuum Cherenkov radiation is then forbidden 
and cannot be used to set experimental limits.  
However, 
the kinematics of
the electromagnetic vertex now allows photon decay 
into fermion--antifermion pairs~\rf{phdec}.  
The dispersion relation~\rf{disprel} remains valid 
and establishes that photons with energies
\beq{gammathres} 
E_{\rm pair}=\frac{2m}{\sqrt{\ktr(\ktr-2)}}
=\sqrt{\frac{2}{-\ktr}}\; m+{\cal O}\left(\sqrt{\ktr}\right) 
\eeq 
or above are unstable, 
where $m$,  as before, 
is the fermion mass.  
In Appendix~\ref{decayrate}, 
we derive the corresponding tree-level decay rate~\rf{exactrate}, 
which to leading order in $\ktr$ is
\beq{pairrate} 
\Gamma_{\rm pair}=\frac{2}{3}\,\alpha\,E_\gamma\,\frac{m^2}{E_{\rm pair}^{2}}\,
\sqrt{1-\frac{E_{\rm pair}^{2}}{E_\gamma^{2}}} \left(2+\frac{E_{\rm
      pair}^{2}}{E_\gamma^{2}}\right)\;.  
\eeq 
Here, 
$E_\gamma$ denotes the photon energy 
and $\alpha$ is again the fine-structure constant~\cite{ks08}.  
The efficiency of this photon decay 
can be established by example: 
a $40\,$GeV photon 
with energy 1\% above threshold 
would decay after traveling an average distance of about $15\,\mu$m.

The above results show 
that we may obtain limits on negative values of $\ktr$ 
from the existence of high-energy long-lived photons.  
As for the Cherenkov analysis, 
cosmic-ray observations provide the potential
to reach the highest sensitivity.  
For example, 
primary photons from the Crab nebula 
with energies up to $80\,$TeV 
have been reported by HEGRA~\cite{HEGRA04}.  
Equation~\rf{gammathres} then implies the possibility 
of one-sided limits on $\tilde{\kappa}$ coefficients 
at the $10^{-16}$ level.  
In addition to some of the nonbirefringent $\tilde{\kappa}$ matrices, 
certain SME coefficients of the electron
cannot be neglected at these scales.  
In view of the small event sample for TeV gamma rays, 
the extraction of comprehensive and clean bounds 
on this potentially large number of SME coefficients 
appears unlikely at the present time.  
Estimates for limits from astrophysical TeV gamma rays 
would certainly be useful,
but such an analysis lies outside the scope of the present study.

We focus instead on high-energy photons generated in Earth-based
laboratories.  As for vacuum Cherenkov radiation, the superior
experimental control allows cleaner, more conservative limits albeit with
reduced sensitivity.  In this context, LEP becomes again one possible
experiment to consider: e.g., past studies of quark-to-photon fragmentation
involved a careful analysis of final-state photons in LEP scattering
events.  In such analyses, photons with energies up to $42\,$GeV have
been studied at OPAL~\cite{OPAL98}, implying such photons survived long enough to interact normally with the OPAL calorimeter.  Equation~\rf{gammathres} then
allows limits at the level $-3\times10^{-10}\lsim\ktr$.  Other LEP
studies at L3~\cite{L302} and OPAL~\cite{OPAL03} detectors, which were
optimized for QED precision tests, have measured pair annihilation
$e^+e^-\to\gamma\gamma$ at center-of-mass energies up to $209\,$GeV.
This would yield an even better bound of $-5\times10^{-11}\lsim\ktr$.

Nevertheless, the highest energies at terrestrial accelerators are not
reached with electrons but with hadrons.  For example, Fermilab's
Tevatron $p\overline{p}$ collider produces center-of-mass energies up
to $1.96\,$TeV and offers therefore excellent potential for
producing high-energy photons.  One particular process, namely
isolated-photon production with an associated jet, is of importance
for QCD studies and has therefore been investigated with the D0
detector.  In this context, photons of energies up to $442\;$GeV have
been observed~\cite{D006}.  The implied stability of photons with such energies   suggests an estimate of
$-3\times10^{-12}\lsim\ktr$.  However, the small number of events
observed at this energy did not warrant inclusion into these QCD
investigations.

Our present analysis uses only D0 photon data at lower energies, where
comparisons to QCD predictions were made.  With this conservative
restriction, photon-energy bins up to $340.5\,$GeV were
measured~\cite{D008}.  For this data, the aforementioned
jet-plus-photon production was measured as a function of $E_\gamma$ in
four angular regions.  These four directional configurations were
characterized by the photon and jet pseudorapidities $y^\gamma$ and
$y^{\rm jet}$.  The largest deviations between experiment and QCD
theory in the $340.5\,$GeV energy bin occurred in the $\{|y^{\rm
  jet}|<0.8, y^\gamma y^{\rm jet}<0 \}$ angular region~\cite{D008}.
The measured cross section was about 52\% of the QCD prediction.  The
relative uncertainties in the experimental value were 46.1\%
statistical, 12.9\% systematic, and a 7.8\% normalization
error~\cite{D008}.  To account for uncertainties, the employed
theoretical scales were varied by a factor of 2, which led to a
relative spread of about 11\% for theoretical predictions~\cite{D008}.
Combining these errors in quadrature yields an overall relative
uncertainty of about 50\%.  The experiment-to-theory ratio in the
$340.5\,$GeV energy bin is therefore $0.52\pm0.26$ for the selected
angular configuration.  We can thus estimate that at least 26\% of the
produced photons have reached the detector.

The layout of the D0 detector implies 
that measured photons 
traverse a minimum distance of $l_{\rm min}\simeq78\,$cm:
they have to travel through 
various drift chambers and the transition-radiation detector
before they interact and are detected 
in the central calorimeter~\cite{D0detector}. 
With the above photon-flux estimate, 
we then obtain
\beq{decaysuppression} 
\exp\left(-\Gamma_{\rm pair}\,l_{\rm min}\right)\geq0.26\;.
\eeq
The $340.5\,$GeV energy bin extended 
from $300\,$GeV to $400\,$GeV. 
We therefore conservatively take $E_\gamma=300\,$ GeV 
in  our analysis.   With Eq.~\rf{pairrate}, 
we then find that $E_{\rm pair}$ 
cannot be more than about $0.1\,$keV
below $E_\gamma$. 
Explicitly including the contribution of $c_e^{00}$, 
we therefore conclude that
\beq{decaybound} 
-5.8\times10^{-12}\leq\ktr-\tfrac{4}{3}c_{e}^{00}\;. 
\eeq  
The uncertainty in the constraint~\rf{decaybound} is essentially
determined by the accuracy of the photon-energy measurement.  As with
the Cherenkov bound, the limit~\rf{decaybound} is larger than the
scale ${\cal S}$, so other photon- or electron-sector coefficients are
not further constrained by this argument.  At the same time, this
justifies the exclusion of these additional coefficients from our
study.

\section{Summary and outlook}
\label{sum}

In this paper, 
we have considered new physical effects arising from
a Lorentz-violating CPT-even deviation of the phase speed of light $c_{\rm ph}$
from its conventional value $c$. 
At the theoretical level, 
such a deviation is controlled by the $\ktr$ coefficient of the SME. 
This coefficient is defined 
with respect to the Sun-centered celestial equatorial coordinate system, 
in which the phase-speed deviation is isotropic. 
At the phenomenological level, 
a positive value for $\ktr$ 
would lead to vacuum Cherenkov radiation~\rf{vcrrec} 
at the rate~\rf{vcrate}
for charges with energies above the threshold~\rf{vcrenergy}; 
whereas a negative value would cause photon-decay~\rf{phdec} 
at the rate~\rf{pairrate} 
for photons with energies above the threshold~\rf{gammathres}.

We have exploited the fact 
that both phenomena 
are efficient threshold effects 
to extract constraints on $\ktr$ 
from the nonobservation of 
vacuum Cherenkov radiation and photon decay. 
In particular, 
the absence of the Cherenkov effect at LEP 
leads to the bound~\rf{VCRbound}, 
and from the stability of photons at the Tevatron 
the constraint~\rf{decaybound} can be inferred. 
These results give the combined conservative limit
\beq{bound} 
-5.8\times10^{-12}\leq\ktr-\tfrac{4}{3}c_{e}^{00}\leq1.2\times10^{-11}\;. 
\eeq
This limit represents an improvement 
of previous laboratory bounds 
by 3--4 order of magnitude.

There are various ways for complementary or improved bounds to be set on $\ktr$.
For instance, planned low-energy laboratory tests could reach a level of $10^{-11}$ or better~\cite{future}.
Another idea is to exploit photon triple splitting, as it is known
that the amplitude for this effect is nonzero in the presence of
$c^{\mu\nu}$ Lorentz violation~\cite{triple}.  This effect does not
involve a threshold, and so high energies are not necessarily
required.

Other future terrestrial bounds could  employ the absence of vacuum Cherenkov
radiation and photon decay at even higher energies than the ones
considered here.  One example would be the prospective International
Linear Collider.  If we take the laboratory-frame energy to be
$500\,$GeV, the International
Linear Collider gives a projected one-sided Cherenkov limit of
$0\leq\ktr-(4/3)c_{e}^{00}\leq5.2\times10^{-13}$.  Similarly, the Large Hadron
Collider will reach about 7 times the energy of the
Tevatron.  Under the assumption that the energy of produced photons
scales by the same factor, the limit~\rf{decaybound} can be tightened
by a factor of 50.  Other improvements of the photon-decay bound would
be possible with a dedicated D0 (or possibly Large Hadron Collider) analysis:
Ultrahigh-energy events not considered for QCD tests could be used
because the statistics of such events is not of primary importance for
photon-decay studies.  Moreover, the end of the photon-energy spectrum
could be exploited more efficiently by avoiding large energy bins.

We note that during the preparation of this manuscript, Brett Altschul performed a more detailed analysis of synchrotron radiation processes at LEP, obtaining an improved two-sided limit on isotropic violations of Lorentz symmetry for light relative to electrons of $|\ktr-(4/3)c_e^{00}|\leq 1.2\times 10^{-15}$~\cite{Altschul:2009b}.

The largest potential for improved bounds 
on $\tilde{\kappa}$ coefficients---in 
the context of both vacuum Cherenkov radiation and photon decay---lies
probably in UHECR physics~\cite{kappatr,ks08}: e.g., with 
a more reliable identification of the UHECR primary particle, 
observations at still higher energies,  better coverage of the sky with more events, 
and data analysis allowing for Lorentz violation in both the primary 
and the decay products
should open an avenue 
to tap this potential more completely. 
A fundamental limit on the experimental reach is that the Universe becomes opaque to cosmic rays 
above certain thresholds
due to processes such as 
Greisen-Zatsepin-Kuzmin suppression or scattering from IR photons.

\acknowledgments 
R.L.~is grateful to B.~Altschul, F.R.~Klinkhamer, and M.~Schreck 
for helpful discussions. 
This work is supported in part by 
the European Commission 
under Grant No.\ MOIF-CT-2005-008687,
by CONACyT under Grant No.\ 55310, 
and by the National Science Foundation.

\appendix

\section{Coordinate rescalings}
\label{rescaling}

In this appendix,  
we present the details of our determination of the scale
$\mathcal{S}$ defined in Eq.~\rf{scale}
at which Lorentz violation has been constrained for
photons relative to electrons, based upon limits obtained from
terrestrial tests.  To accomplish this, we compare the results
of experiments reporting bounds on Lorentz violation for
photons, electrons, and protons.  Making use of these bounds is
somewhat complicated by the assumptions made regarding the possibility
of Lorentz violation in other sectors of the SME.

Measurements of shifts and anisotropies in the vacuum speed of light
must be defined in terms of the velocity of a chosen reference
particle.  In the SME, the limiting velocity of any such reference
particle is also subject to Lorentz-violating shifts and
anisotropies.  As a result, constraints upon the deviation of the
speed of light based upon interactions of light with electrons must be
narrowly interpreted as limits on the difference between the degree to
which Lorentz symmetry is violated in each sector.  This is a general
feature of all tests of Lorentz symmetry, which generally must be
described as setting limits on combinations of coefficients associated
with not one, but all involved particle species.  
In this sense, 
the number 
of independent SME coefficients 
controlling the type and extent of Lorentz violation
in a given physical system
is {\em increased}.

Under certain circumstances, 
various SME coefficients describing a given system 
may be physically equivalent 
and can therefore not be distinguished. 
From a mathematical viewpoint, 
there typically exist canonical transformations 
that can eliminate one coefficient in favor of the other. 
From an experimental viewpoint, 
this means that such coefficients 
cannot be bounded or measured independently
in the physical system in question. 
In such a case, 
the number of independent SME coefficients 
is therefore effectively {\em reduced}. 

It is the interplay of the above two issues
that is often unappreciated in the literature. 
The interpretation of experimental constraints 
therefore requires special care. 
In the present context, 
the former issue has been discussed 
in various places in main text. 
The latter issue is not only paramount 
for the precise formulation of our actual bounds,
but also for the interpretation of existing constraints 
necessary for the determination of the scale~$\mathcal{S}$. 
The particular issue 
to be clarified in this appendix 
concerns the $\tilde{k}^{\mu\nu}$ coefficient 
(i.e., $\kem$, $\kop$, $\ktr$) 
and $c^{\mu\nu}$-type coefficients. 
To simplify the discussion, 
we will set all other types of SME coefficients 
to zero in what follows.

In the context of the SME for one-flavor QED, a theory characterized by a nonzero 
symmetric, traceless $c_{e}^{\mu\nu}$ tensor and a vanishing $(k_{F})^{\kappa\lambda\mu\nu}$ tensor
exhibits the same phenomenology as a theory in which
$c_{e}^{\mu\nu}=0$~\cite{km0102,bk04} and
\beq{specialk}
(k_{F})^{\mu\nu\rho\sigma}=
\frac{1}{2}(\eta^{\mu\rho}\tilde{k}^{\nu\sigma}
-\eta^{\mu\sigma}\tilde{k}^{\nu\rho}
+\eta^{\nu\sigma}\tilde{k}^{\mu\rho}
-\eta^{\nu\rho}\tilde{k}^{\mu\sigma}),
\eeq
provided that 
\beq{equivalence}
\tilde{k}^{\mu\nu}=-2c_e^{\mu\nu}
\eeq
at linear order. 
Thus, 
a $\tilde{k}^{\mu\nu}$ model is
physically equivalent to a $c_e^{\mu\nu}$ model
if the models are related by Eq.~\rf{equivalence}.
This fact can also be formally
established via coordinate rescalings~\cite{km0102,bk04}.  
We may use this freedom 
to select a particularly convenient 
scaling of the coordinates 
to simplify calculations. 
For example, 
our analysis in the main text 
is performed within a $\tilde{k}^{\mu\nu}$ model
(i.e., the coordinates are scaled such that $c_e^{\mu\nu}=0$),
whereas our photon-decay calculation in Appendix~\ref{decayrate}
employs a $c_e^{\mu\nu}$ model
(i.e., the coordinates are rescaled such that $\tilde{k}^{\mu\nu}=0$). 
One way of quoting results,
such as experimental constraints, 
is to employ a particular coordinate scaling
and {\em clearly state this special scaling choice} 
together with the actual result.

The choices of rescaling form a continuous set 
and are not only confined to the two canonical cases 
of $c_e^{\mu\nu}=0$ and $\tilde{k}^{\mu\nu}=0$ 
discussed above. 
An infinite number of coordinate scalings with both  
$c_e^{\mu\nu}\neq0$ and $\tilde{k}^{\mu\nu}\neq0$ 
can certainly be selected. 
In the present context, 
one can show 
that with such general rescalings, 
physical effects can only depend upon (and thus
provide bounds for) the value of $2c_{e}^{\mu\nu}-\tilde{k}^{\mu\nu}$. 
This is intuitively reasonable 
because the two Lorentz-violating effects considered in this work,
vacuum Cherenkov radiation and photon decay, 
depend only on certain velocity {\em differences} 
between the electron and the photon.
Note in particular
that the combination $2c_{e}^{\mu\nu}-\tilde{k}^{\mu\nu}$ 
does not 
pertain to a particular choice of scale for the coordinates
and therefore provides a second way
to quote results
that is {\em coordinate-scaling independent}. 
When formulating our final bounds in the main text,
we have adopted this latter choice of stating results. 

Other experimental tests of Lorentz and CPT symmetry 
are not generally confined to
one-flavor QED; other particle species are often involved.  In such
situations, the above analysis is readily generalized: we may choose
one particle species to serve as the reference ``ruler,'' and thus
work in a coordinate system in which that species' $c^{\mu\nu}$
coefficient (or $\tilde{k}^{\mu\nu}$, if the reference is light) is
zero.

With these considerations, 
we estimate the value $\mathcal{S}$ provided by
terrestrial experiments for the interaction of light with electrons.
Specifically, we will use the results of a Cs-fountain clock
experiment~\cite{wolf06}, and those of a series of tests involving
optical resonators~\cite{her07}.  These tests are sensitive to Lorentz
violation in conventional matter, which is made up of protons,
neutrons, and electrons, interacting electromagnetically.  In this four-species systems, the Lorentz-violating effects under consideration are described by four sets of SME coefficients: $c_{p}^{\mu\nu}$, $c_{n}^{\mu\nu}$, $c_{e}^{\mu\nu}$, and $\tilde{k}^{\mu\nu}$, where the subscripts $p$, $n$, and $e$
respectively denote the coefficients of protons, neutrons, and electrons.  One of these four sets of terms may be eliminated by a scaling of coordinates, and so in practice only three of these sets of parameters may independently contribute to the physics.

Although the Cs-fountain test~\cite{wolf06} involves protons,
neutrons, electrons, and electromagnetism, the observed frequencies
turn out to be only sensitive to the value of
$2c_{p}^{\mu\nu}-\tilde{k}^{\mu\nu}$.  The constraints upon the eight
spatially anisotropic components are at the level of
\beq{Cs-fountain}
|2c_p^{\mu\nu}-\tilde{k}^{\mu\nu}|<10^{-21}\ldots10^{-25}\;.
\eeq
We note that these results are presented in Ref.~\cite{wolf06} in
coordinates such that $\tilde{k}^{\mu\nu}=0$, which corresponds to
using light as a reference.  In coordinates with protons as the reference (i.e., $c^{\mu\nu}_p=0$), 
the Cs-fountain
experiment provides the constraint
$|\tilde{k}^{\mu\nu}|<10^{-21}\ldots 10^{-25}$ for the anisotropic
components of $\tilde{k}^{\mu\nu}$.

Next, we consider optical-resonator experiments~\cite{her07}, which
measure the resonance frequencies $\nu$ of light propagating in vacuum
inside two orthogonally oriented Fabry-P\'erot cavities.  As
previously shown~\cite{kr,photonexpt}, these experiments are
sensitive to spatial anisotropies in the speed of light
($\tilde{k}^{\mu\nu}$) and to variations in the dimensions of the
resonators themselves.  The cavity size is primarily determined by the
electromagnetic interactions in the chemical bonds.  It therefore
follows that the neutron's contribution to the cavity size must be
suppressed because it is uncharged, having only a magnetic moment.
Moreover, the cavities are made of fused silica SiO$_{2}$, and the
common isotopes of oxygen and silicon have even numbers of neutrons
and spin zero.  Pairing effects would therefore tend to further
suppress the influence of the neutron spin, and so we conclude that
the cavity frequencies should be largely unaffected by
$c_{n}^{\mu\nu}$.  

Bearing this suppression of neutron effects in mind,
two independent combinations of parameters remain
that can influence the observable $F$ determined from the cavity
frequencies $\nu$.  
This observable must therefore be given, to leading order, 
by an expression of the form
\beq{vacfreq}
F=\textrm{const.}+A_{\mu\nu}(2c_e^{\mu\nu}-\tilde{k}^{\mu\nu})
+B_{\mu\nu}(2c_{p}^{\mu\nu}-\tilde{k}^{\mu\nu})\;,
\eeq
where $A_{\mu\nu}$ and $B_{\mu\nu}$ are constants. 
Constraints on the anisotropic pieces in Eq.~\rf{vacfreq} 
at the $10^{-13}\ldots 10^{-17}$ level
can be obtained by these cavity tests~\cite{her07}.

The $B_{\mu\nu}$ term in Eq.~\rf{vacfreq}
can be dropped from these optical-resonator bounds 
for the following reason: 
The constants $A_{\mu\nu}$ and $B_{\mu\nu}$ 
are likely to be of similar size,
as there appears to be no convincing argument 
suggesting that the SME effects in one of the three involved particle species 
would dominate the length of chemical bonds. 
For certain sample chemical bonds including fused silica, 
this has indeed been verified~\cite{bonds}. 
The next step is to observe 
that the independent Cs-fountain bound~\rf{Cs-fountain}
places a much tighter constrained 
on the coefficient combination multiplying $B_{\mu\nu}$ 
than the best sensitivity $10^{-17}$ of the optical-resonator test.
We thus conclude that
\beq{kappabounds}
|2c_e^{\mu\nu}-\tilde{k}^{\mu\nu}|<10^{-13}\ldots 10^{-17}
\eeq
follows from the cavity experiments~\cite{her07}.
It is understood 
that this bound refers to the anisotropic components of $2c_e^{\mu\nu}-\tilde{k}^{\mu\nu}$. 
We remark that Ref.~\cite{her07} chooses to state 
the resulting experimental limit
assuming scaled coordinates such that $c_e^{\mu\nu}=0$. 

The constraint~\rf{kappabounds} 
taken together with Refs.~\cite{kr,eexpt2and3} establish 
that $\mathcal{S}\sim 10^{-13}$, dominated by the contribution of the
parity-odd $\kop$ and $c_{e}^{0J}$ coefficients.  
Because the limit we will derive on isotropic $\ktr$ component 
lies above this scale, 
we may indeed drop all other
Lorentz-violating corrections from our analysis.

\section{Photon-decay rate}
\label{decayrate} 

Photon-decay rates 
in the presence of Lorentz violation 
have been determined~\cite{gammadecay} for the
 dimension-three Chern-Simons type 
SME coefficient $(k_{AF})^{\mu}$ which governs photon triple splitting.
In this appendix, 
we derive the tree-level photon-decay rate 
into a fermion--antifermion pair
arising from the dimension-four SME $\ktr$ coefficient, 
appropriate for our purposes.

The starting point is a model 
with Lorentz-violating photons 
and conventional charged leptons. 
In the present situation 
it is convenient 
to consider a physically equivalent model
constructed with the coordinate redefinition discussed in Sec.~\ref{review}
and Appendix~\ref{rescaling}. 
In particular, 
we remove all Lorentz violation from the photon sector 
at the cost of introducing a Lorentz-breaking $c_e^{\mu\nu}$ coefficient
in the lepton sector: 
\beq{lvqed}
{\cal L}'=\half i\; \overline{\hskip-2pt\psi}\,(\gamma^\mu+c_e^{\mu\nu}\gamma_\nu)\lrDmu\psi
-m\; \overline{\hskip-2pt\psi}\psi -\quar F^2\;,
\eeq
where $D_\mu=\prt_\mu+ieA_\mu$ is the usual covariant derivative. 
The Lorentz-violating SME coefficient $c_e^{\mu\nu}$ is given in explicit form as
\beq{equivalence2} 
c_e^{\mu\nu}=-\fr{1}{4}\ktr\,\textrm{diag}(3,1,1,1). 
\eeq 
The advantage of the above description \rf{lvqed} for $\ktr$ Lorentz violation 
is the following. 
Perturbation theory in quantum field theory 
relies on the quantization of the free-field sectors of the model. 
For Lorentz-violating photons, 
such a quantization is lacking, 
but the quantization of SME fermions 
is comparatively well understood~\cite{sme}. 
Moreover, 
we may employ the methodology 
and notation
of a previous tree-level calculation 
involving $c_e^{\mu\nu}$ fermions~\cite{scattering}.

Since $\ktr-(4/3)c_{e}^{00}$ is nonzero, 
the lepton sector of the model~\rf{lvqed} in our chosen coordinates
contains unconventional time derivatives. 
The time evolution of $\psi$ 
can then be nonunitary, 
so its asymptotic states cannot directly be identified 
with physical free-particle states. 
A standard approach 
to avoid this potential interpretational difficulty 
is a redefinition of the spinor field 
chosen to eliminate the additional time derivatives~\cite{scattering,rl04}. 
In the present situation,
the field redefinition amounts to a rescaling
\beq{fredef}
\psi\equiv\frac{1}{\sqrt{1-\frac{3}{4}\ktr}}\chi\;,
\eeq
so it is not strictly mandatory. 
We nevertheless implement the redefinition~\rf{fredef} 
 for compatibility with previous studies~\cite{scattering} of this type.
The Lagrangian~\rf{lvqed} then becomes
\beq{lvqed2}
{\cal L}=\half i\; \overline{\hskip-2pt\chi}\;\tilde{\eta}^{\mu\nu}\gamma_\nu\lrDmu\chi
-\tilde{m}\; \overline{\hskip-2pt\chi}\chi -\quar F^2\;,
\eeq
where we have defined
\bea{tildedefs}
\tilde{m} & \equiv & \frac{m}{1-\frac{3}{4}\ktr}\;,\nonumber\\
\tilde{\eta}^{\mu\nu} & \equiv & \textrm{diag}(1,-\lambda,-\lambda,-\lambda)\;,\nonumber\\
\lambda & \equiv & \frac{1+\frac{1}{4}\ktr}{1-\frac{3}{4}\ktr}\;.
\eea
The field redefinition~\rf{fredef} 
is a canonical transformation, 
and as such it leaves unchanged the physics.
In particular, 
the free fermions in model~\rf{lvqed} 
possess the same dispersion relation
as those in \rf{lvqed2}. 

Because $\ktr$ is frame dependent, Eq.~\rf{fredef} is not manifestly
Lorentz coordinate covariant. Therefore, the specific form of
Lagrangian~\rf{lvqed2} holds only in the frame in which the field
redefinition has been performed \cite{fn2}.  Note that the
Lagrangians~\rf{lvqed} and~\rf{lvqed2} are singular for $\ktr=-4$ and
$\ktr=4/3$ while the dispersion relation~\rf{disprel} is
singular at $\ktr=1$.  This difference arises because the coordinate
rescaling used to generate Lagrangian~\rf{lvqed} from the original
$\ktr$ model has been implemented only at leading order in $\ktr$.

\begin{figure}[t]
\vskip10pt
\begin{center}
\includegraphics[width=0.60\hsize]{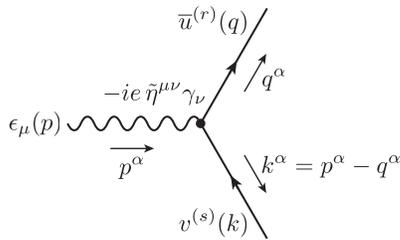}
\end{center}
\vskip-10pt
\caption{Tree-level Feynman diagram for photon decay. 
Lorentz-violating effects are contained 
in the modified dispersion relation 
for the lepton and antilepton 4-momenta 
$q^\alpha$ and $k^\alpha$, respectively,
as well as in the electromagnetic vertex 
containing $\tilde{\eta}_{\mu\nu}$.}
\label{fig1} 
\end{figure} 

The Feynman rules can now be inferred from the Lagrangian~\rf{lvqed}.
The appropriate tree-level Feynman diagram for photon decay is
depicted in Fig.~\ref{fig1}.  For the corresponding matrix element, we
obtain
\beq{melement} 
i{\cal M}_{rs}=-ie\,\epsilon_{\mu}(p)\,\tilde{\eta}^{\mu\nu} 
\,\overline{u}^{(r)}(q)\,\gamma_\nu\, v^{(s)}(k)\;,
\eeq
where the various polarization and momentum assignments 
are defined in Fig.~\ref{fig1}. 
The next step is the calculation of $|{\cal M}_{rs}|^2$ 
followed by the usual summation over final spin states
and averaging over the initial photon polarizations
$\overline{|{\cal M}|^2}\equiv(1/2)\sum_\epsilon\sum_{r,s}|{\cal M}_{rs}|^2$. 
We obtain
\beq{prob}
\overline{|{\cal M}|^2}=e^2\!
\left[4\tilde{m}^2+2\lambda^2(1-\lambda^2)(\vec{q}^{\,2}+\vec{k}^{2})
+(1-\lambda^2)^2E_\gamma^2\right],
\eeq
where $E_\gamma=|\vec{p}|$ is the photon energy
and $\vec{q}$ and $\vec{k}$ are the lepton and antilepton 3-momenta, 
respectively. 
To arrive at this result, 
energy--momentum conservation, 
the usual relation for photon-polarization sums, 
and trace identities for Dirac matrices
have been used. 
Moreover, 
we have employed the results 
for SME spinor projectors in Ref.~\cite{scattering} 
with the normalization chosen 
such that $N(\vec{q})=2E_q=2\sqrt{\tilde{m}^2+\lambda^2\vec{q}^{\,2}}$, etc.

The final step is the phase-space integration. 
In the conventional Lorentz-symmetric case, 
the decay rate for massive particles is defined 
in the particle's rest frame 
with a kinematic factor inversely proportional to its mass. 
This procedure cannot be applied
to present massless case.
We adopt instead the convention~\cite{triple}
to define the decay rate 
in terms of the photon energy $E_\gamma$ 
in the Sun-centered celestial equatorial frame:
\beq{ratedef} 
\Gamma_{\rm pair}=\frac{1}{4\pi^2}\frac{1}{2E_\gamma}
\int\!\frac{d^3q}{2E_q}\frac{d^3k}{2E_k}
\,\overline{|{\cal M}|^2}\,\delta^{(4)}(p-q-k)\;.
\eeq
This yields
\beq{exactrate}
\Gamma_{\rm pair}=\alpha\frac{[\ktr(\ktr-4) E_\gamma^2+4m^2]
\sqrt{\ktr(\ktr-4) E_\gamma^2-8m^2}}
{\frac{3}{8}E_\gamma^2(4+\ktr)(4-3\ktr)\sqrt{\ktr(\ktr-4)}}
\eeq
for the exact tree-level decay rate within the context of
Lagrangian~\rf{lvqed2}.  Equation~\rf{exactrate} applies only for
perturbative $\ktr<0$ and for photons above threshold.  We remark that
undoing our initial coordinate redefinition would generate subleading
corrections to Eq.~\rf{exactrate}.


\begin{thebibliography}{99}
\bibitem{fn0}
The analysis in Ref.~\cite{kappatr} implicitly 
scales the coordinates such that $c_{p}^{\mu\nu}=0$.
\bibitem{fn0.5}
In fact, 
in Appendix~\ref{rescaling} we recall 
that tests with Cs-fountain clocks 
bound 
the eight anisotropic components 
of $2c_{p}^{\mu\nu}-\tilde{k}^{\mu\nu}$ 
with sensitivities of $10^{-21}\ldots 10^{-25}$, 
so that some of these eight components 
can actually be dropped from the UHECR analysis 
in Ref.~\cite{kappatr}.
\bibitem{fn1} This does not necessarily entail violations of
  causality~\cite{caus}.  For example, we can reinterpret the model
  employing the coordinate rescaling discussed in Sec.~\ref{review}.
  In the rescaled Minkowski spacetime, photons moves on the new light
  cone; fermions, which now exhibit $c^{\mu\nu}$ Lorentz violation,
  have a maximal speed less than that of the photon.  Thus, both the
  photon and the fermion obey conventional causality with respect to
  the new spacetime.
\bibitem{fn2}
In the present case, 
we have selected 
the Sun-centered celestial equatorial frame. 
Since the $\ktr$ term is isotropic, 
Lagrangian \rf{lvqed2} also holds in all inertial Minkowski coordinate systems 
at rest relative to the Sun-centered celestial equatorial frame.
\end{thebibliography}
\end{document}